\documentclass[12pt]{article}
\usepackage{graphicx}
\usepackage{type1cm, amsmath,hangcaption}
\usepackage[psamsfonts]{amssymb}
\usepackage{cite}
\numberwithin{equation}{section}

\renewcommand{\caption}{\isucaption}

\newcommand{\Ncal}{{\cal N}}
\newcommand{\del}{\partial}
\newcommand{\vp}{\varphi}
\newcommand{\gf}{$g$}
\newcommand{\nn}{\nonumber}
\newcommand{\Ib}{I_{\text{brane}}}

\begin{document}
%%%%%%%%%%%%%%%%%%%% title page %%%%%%%%%%%%%%%%%%%%%
\thispagestyle{empty}
\begin{flushright}
 \parbox{3.5cm}{UT-02-39 \\
  {\tt hep-th/0207171}}
\end{flushright}

\vspace*{2cm}
\begin{center}
 {\Large
 %%%  title %%%
    Holographic RG Flow on the Defect and \gf-Theorem
 }
\end{center}

\vspace*{3cm} 
%%%%%%  author  %%%%%%
\begin{center}
 \noindent
 {\large Satoshi Yamaguchi}

 \vspace{5mm}
 \noindent
 \hspace{0.7cm} \parbox{130mm}{\it
 Department of Physics, Faculty of Science, University of Tokyo,\\
 Tokyo 113-0033, Japan.\\
 E-mail: {\tt yamaguch@hep-th.phys.s.u-tokyo.ac.jp}
}
\end{center}

\vspace{3cm}
\hfill{\bf Abstract\ \ }\hfill\ \\
%%%%%%%%%%%%%%%%%%%%%%%%%   abstract   %%%%%%%%%%%%%%%%%%%%%%%
We investigate relevant deformation and the renormalization group flow
in a defect conformal field theory from the point of view of the holography.  We 
propose a candidate of \gf-function in the context of the holography, and prove the \gf-theorem: the \gf-function is monotonically non-increasing along the RG flow.
We apply this \gf-theorem to the
 D5-brane solution which is an asymptotically $AdS_4 \times S^2$ brane in 
$AdS_5\times S^5$. This solution corresponds to the mass
deformation of the defect CFT. We checked that the \gf-function is monotonically 
non-increasing in this solution.

\newpage
%%%%%%%%%%%%%%%%%%%% end of title page %%%%%%%%%%%%%

\section{Introduction}

Quantum field theories with defects (domain walls) appear in various fields of
 physics. This class of theories includes boundary conformal field theory, and
 provide the excellent worldsheet description of D-branes 
\cite{Polchinski:1995mt}.
 Defect quantum field theories are also useful in impurity
 problem in condensed matter physics 
\cite{McAvity:1995zd,Oshikawa:1997dj,Saleur:1998hq}.

Some defect conformal field theories have been considered
\cite{Karch:2000gx,DeWolfe:2001pq,Bachas:2001vj,Erdmenger:2002ex,%
   Karch:2002sh,Mateos:2002bu}
in the framework of AdS/CFT correspondence\cite{Maldacena:1998re,Aharony:1999ti}.
In this context, branes on AdS correspond to defects in CFT. The defects which 
do not break the scale invariance correspond to AdS shaped branes.
In $AdS_5\times S^5$ for example, a D5-brane 
wrapped on $AdS_4 \times S^2$ corresponds to a defect which preserves
the scale invariance of 4-dimensional $\Ncal=4$ super Yang-Mills theory.

In ordinary AdS/CFT correspondence, some relevant deformation and 
renormalization group flow (holographic RG flow) have been considered 
by the deformation of the geometry of 
AdS\cite{Henningson:1998gx,Alvarez:1998wr,Girardello:1998pd,%
Balasubramanian:1999jd,%
Freedman:1999gp,Girardello:1999bd,Skenderis:1999mm,%
deBoer:1999xf,Nojiri:1999eg,Fukuma:2000bz}. 
The coupling constants for relevant operators in field 
theory correspond to tachyonic fields (satisfying Breitenlohner-Freedman bound 
\cite{Breitenlohner:1982bm}) in AdS, and energy scale corresponds to the radial 
coordinate of AdS. Especially, the c-function is defined in gravity side
and proven to be monotonically non-increasing along the flow.

It is also essential to consider relevant deformation on the defect and the
renormalization group flow. We investigate this relevant deformation on defect 
in the framework of the holography. A solution of an flow has been constructed in 
\cite{Skenderis:2002vf}. Another type of solution is obtained in \cite{Karch:2002sh}.

First, we consider the mass deformation of
the fields on the defect. We consider the deformation of 
the $AdS_4\times S^2$ D5-brane in $AdS_5\times S^5$, whose holography to defect 
conformal field theory is considered in \cite{DeWolfe:2001pq}. We explain the 
solution of the field theory on the brane \cite{Karch:2002sh}.
The IR theory on the defect turns out to be massive as expected. This means
that the defect decouples from the ambient theory and vanishes in IR.

Next, we propose a candidate of the \gf-function in the holography, and prove 
the \gf-theorem. \gf-function (boundary 
entropy)\cite{Affleck:1991tk} in defect field theory is an analogue of 
c-function in ambient field theory. It has been conjectured  
in \cite{Affleck:1991tk} that the 
\gf-function is monotonically non-increasing in the flow (\gf-theorem), which 
mean $g_{\rm UV}\ge g_{\rm IR}$.

Let us note here about the terminology. The words ``boundary'' and ``bulk'' are 
very confusing in this context. Therefore we use the following words in this 
paper. We call each side of the duality ``gauge theory'' and ``string theory''. 
As for the gauge theory side, we use ``defect'' and ``ambient''. In the string 
theory side, we use ``brane'' and ``gravity''.

\section{The solution representing the flow}\label{section-solution}
In this section, we explain a D5-brane
 solution, which represents the mass deformation on the 
defect. This solution has been first obtained by Karch and Katz \cite{Karch:2002sh}
\footnote{
Although we independently obtained the result explained in this section,
after submitted this paper to e-print archive, we become aware
of the earlier paper \cite{Karch:2002sh} in which the same result appeared.
I would like to thank Andreas Karch for letting me know this.
}.
Some features of this solution has been also discussed in \cite{Mateos:2002bu}
\footnote{After submitted this paper to e-print archive, we also become aware
that some result in this section
has been obtained in \cite{Mateos:2002bu}. I would like to
thank David Mateos and Selena Ng for letting me know this.}.

\subsection{Setting}
We consider the near-horizon geometry of $N$ D3-branes. This background is
$AdS_5\times S^5$ with RR 5-form flux. Let us denote the coordinates of
$AdS_5$ by $(\rho,x^{\mu}),\ (\mu=0,1,2,3)$ and the coordinates of $S^5$
by $(\psi,\theta_1,\vp_1,\theta_2,\vp_2)$. The near-horizon metric can be 
written as 
\begin{align}
 ds^2=R^2\left[ d\rho^2 + e^{-2\rho}\eta_{\mu\nu}dx^{\mu}dx^{\nu}
  +d\psi^2+\cos^2{\psi}(d\theta_1^2+\sin^2\theta_1 d\vp_1^2)
  +\sin^2{\psi}(d\theta_2^2+\sin^2\theta_2 d\vp_2^2)\right],
  \label{metric}
\end{align}
where $R=(4\pi\alpha'{}^2 g_s N)^{1/4}$, 
and $\eta_{\mu\nu}={\rm diag}(-1,1,1,1)$.
The RR 5-form flux $F_{(5)}$ is expressed as
\begin{align*}
 F_{(5)}=\frac{4 R^4}{g_s} (e^{-4\rho}dx^{0} dx^{1} dx^{2} dx^{3} d\rho 
          + \cos^2\psi\sin^2\psi \sin \theta_1\sin \theta_2 d\psi d\theta_1 
		   d\vp_1 d\theta_2 d\vp_2).
\end{align*}
The RR 4-form potential reads
\begin{align}
 C_{(4)}=\frac{4R^4}{g_s} (-x^3 e^{-4\rho}dx^{0} dx^{1} dx^{2} d\rho
     + \vp_2 \cos^2\psi\sin^2\psi \sin \theta_1\sin \theta_2 d\psi d\theta_1
       d\vp_1 d\theta_2).
   \label{4-form}
\end{align}

We explore a single D5-brane in this background by considering the theory on 
D5-brane: Dirac-Born-Infeld term $+$ Chern-Simons term. In our convention, the 
bosonic part of this D5-brane action in this background becomes
\begin{align}
 I_{5}=-\frac{\mu_5}{g_s} \int d^6 \xi \sqrt{-\det(G_{ab}+2\pi \alpha' F_{ab})}
      -\frac{\mu_3}{2\pi} \int F\wedge C_{(4)}, \qquad a,b=0,\dots,5,
 \label{brane-action1}
\end{align}
where $\mu_{p}:=(2\pi)^{-p}(\alpha')^{-(p+1)/2}$ ,  $F_{ab}$ is the U(1) gauge 
field strength on the brane, $G_{ab}$ is the pullback of the spacetime 
metric (\ref{metric}), and $C_{(4)}$ is pullback of the RR 4-form potential
(\ref{4-form}).

A simple solution of the system (\ref{brane-action1}) is the $AdS_4\times S^2$ 
brane expressed as
\begin{align}
 &x^{j}=\xi^{j},\quad j=0,1,2,\qquad \rho=\xi^{4},\qquad 
 \theta_1=\xi^{5},\qquad \vp_{1}=\xi^{6}, \nn\\
&x^3=\psi=\theta_2=\vp_2=0, \qquad F_{ab}=0. \label{conformal-solution}
\end{align}

Let us make a few comments about this solution. 
First, this solution is the simplest one among the solutions 
constructed in \cite{Karch:2000gx}, and the two ambient theory living in 
left and right of the defect in associated gauge theory are the same theory. 
The associated defect CFT has been investigated in detail in 
\cite{DeWolfe:2001pq}. In this paper, we mainly consider this simplest case. 

Secondly, this solution is BPS as shown in \cite{Skenderis:2002vf}, and has a 
tachyonic mode $\psi$ \cite{Karch:2000gx}, 
which satisfies the  Breitenlohner-Freedman bound \cite{Breitenlohner:1982bm}.
We consider next the deformation of this solution by this tachyonic mode.

\subsection{Mass deformation and D-brane solution representing the flow } 
\label{subsection-solution} 

Now, let us consider the deformation of $AdS_4\times S^2$ brane mentioned in the
previous section. First, we consider the meaning of the field $\psi$ in the 
gauge theory. In
 \cite{DeWolfe:2001pq}, it is claimed from the conformal dimension and the R 
 charge that the operator which correspond to $\psi$ is the mass term 
  on the defect
\begin{align}
 \int d^3 x d^2\Theta m \sigma^{1}_{ij}\bar Q^{i}  Q^{j},\label{mass-term}
\end{align}
where $m$ is the coupling constant (mass) and $\sigma^{1}$ is one of the Pauli
matrices, and $Q^{i},\ \bar Q^{i},\ i=1,2$  are the superfield on the defect
\begin{align*}
 Q^{i}(x, \bar\Theta, \Theta)=q^{i}(x)+ \bar \Theta \Psi^{i}(x) + \frac12 \bar\Theta 
   \Theta f^{i}(x),\\
 \bar Q^{i}(x, \bar\Theta, \Theta)=\bar q^{i}(x)+ \bar \Psi^{i}(x) \Theta 
                + \frac12 \bar\Theta \Theta \bar f^{i}(x),
\end{align*}
and $\Theta, \bar\Theta$ are the fermionic coordinates of the superspace.
The deformation (\ref{mass-term}) is relevant, and the conformal dimension 
shows that the leading term of the flow equation are expressed as 
\begin{align}
 \Lambda \frac{d}{d \Lambda}m= - m +\cdots, \label{flow-equation0}
\end{align}
where $\Lambda$ is the energy cut-off.

Next, let us turn to the string theory side and solve the flow.
Since we deform the theory by the operator on the defect, the bulk theory 
does not flow. This means in string theory side that only the 
$AdS_4\times S^2$ brane is deformed, but 
the background geometry $AdS_5\times S^5$ is not deformed. 
Consequently, the theory governs this flow is 
the one expressed by the brane action (\ref{brane-action1})
\footnote{Since we consider the configuration with 3-dimensional Poincare 
symmetry, the fermion fields do not have vacuum expectation value.}. 

In the region of negative large $\rho$ (UV in gauge theory side), 
we add a small perturbation to $\psi$ (independent of $x^{a},\ a=0,1,2$). Then 
$\psi$ varies along $\rho$ 
according to the action (\ref{brane-action1}). In gauge theory side, 
this motion is renormalization group flow. 
We make $\psi$ dependent on $\xi^4=\rho$, that is, $\psi=\psi(\rho)$, 
and other fields are the same as Eq. (\ref{conformal-solution}). 
Since we want to find a solution 
asymptotically $AdS_4\times S^2$, we impose the boundary condition $\psi\to0$ 
when $\rho\to -\infty$ . 

If we insert this anzats to the D5-brane action (\ref{brane-action1}), the 
action becomes
\begin{align}
 I_5&=-T_5\int d^3 x \int d\rho \int_{0}^{\pi}d\theta_1\int_0^{2\pi} d\vp_1
  R^6 e^{-3\rho} \cos^2\psi \sin\theta_1 \sqrt{1+(\psi')^2}\nn\\
   &=-4\pi R^6 T_5 \int d^3 x \int d\rho e^{-3\rho} \cos^2\psi \sqrt{1+(\psi')^2},
    \label{brane-action2}
\end{align}
where $T_5:=\mu_5/g_s$ is the D5-brane tension, and $\psi':=d\psi/d\rho$. 

We can derive the equation of motion of $\psi$ from the action 
(\ref{brane-action2}). However, in this method, the relation between equation of 
motion and RG flow equation is not clear; the equation of motion is a 
second order differential equation, while the RG flow equation is
a first order one. We can derive the first order differential equation from
the action (\ref{brane-action2}) by 
considering the Bogomolnyi-like bound. In this case, we are considering the 
supersymmetric flow.

To derive Bogomolnyi-like equation, we denote the relevant factor of 
(\ref{brane-action2}) as $I_{\psi}$, and rewrite it as
\begin{align}
 I_{\psi}&:= \int d\rho e^{-3\rho} \cos^2\psi \sqrt{1+(\psi')^2}\nn\\
        &= \int d\rho e^{-3\rho} \cos^2\psi 
        \sqrt{(\cos\psi+ \psi'\sin\psi )^2+(\sin\psi- \psi'\cos \psi )^2}\nn\\
       &\ge \int d\rho e^{-3\rho} \cos^2\psi 
        \sqrt{(\cos\psi+ \psi'\sin\psi )^2} \nn\\
       &\qquad =\frac13\left[e^{-3A}\cos^3\psi(A)-e^{-3B}\cos^3\psi(B)\right].
 \label{BPS-bound}
\end{align}
In the last equality, we denote the range of integration $A\le \rho \le B$ .
The condition to saturate the inequality (\ref{BPS-bound}) leads to the
Bogomolnyi-like equation
\begin{align}
 \sin\psi-\psi'\cos \psi =0. \label{BPS-equation}
\end{align}
It can be checked that if $\psi$ satisfies Eq.(\ref{BPS-equation}), then this
 $\psi$ satisfies the equation of motion derived from Eq.(\ref{brane-action2}).

We claim that Eq.(\ref{BPS-equation}) represents the flow equation in the gauge
 theory and $\psi$ is proportional to $m$. In order to see the asymptotic 
 consistency, we write the relation between the energy cut-off $\Lambda$ and 
 radial coordinate $\rho$. This relation can be read from the coefficient of 
 $\eta_{\mu\nu}dx^{\mu}dx^{\nu}$ in the metric (\ref{metric}) as
\begin{align*}
 \Lambda=\Lambda_0 e^{-\rho},\qquad (\Lambda_0:\text{constant}).
\end{align*}
If we take this relation into account, we can rewrite Eq.(\ref{BPS-equation}) as
\begin{align*}
 \Lambda\frac{d}{d\Lambda}\psi=-\tan\psi=-\psi+\cdots.
\end{align*}
This is consistent with Eq.(\ref{flow-equation0}).

The flow equation (\ref{BPS-equation}) can be easily solved. The solution 
becomes 
\begin{align}
 \psi=\arcsin e^{\rho}, \label{solution}
\end{align}
where we suppress the irrelevant integral constant. The branch of $\arcsin$ is 
taken to $0 \le \psi \le \pi/2$. Let us make a few comments on the solution (\ref{solution}).

First, the solution (\ref{solution}) is only valid in $\rho \le 0$. In $\rho > 
0$, there is no brane. This fact shows that in IR region in the gauge theory, 
the fields on the defect becomes massive, and decouple from ambient. This means, 
in IR region, the defect vanishes and the CFT without defect is realized.

\begin{figure}
\begin{center}
  \includegraphics[width=6cm]{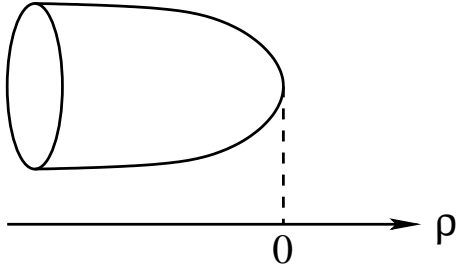}
\end{center}
\caption{The form of the brane. There are no singularity on the brane at 
$\rho\to 0$.}
\label{fig1}
\end{figure}
Secondly, the solution (\ref{solution}) looks singular at $\rho=0$, but this 
brane is 
not singular when we consider the whole brane as illustrated in figure \ref{fig1}.
In order to see this, we write the induced metric on the brane of this solution
\begin{align*}
R^{-2} ds^{2}_{\text{brane}}=e^{-2\rho}\eta_{ij}dx^{i}dx^{j}
     +\left[1+(\psi'(\rho))^2\right]d\rho^2
   +\cos^2\psi(\rho)\left(d\theta_1^2+\sin^2 \theta_1 d\vp_1^2\right).
\end{align*}
In $\rho\to 0$, the first term is not singular, but the second and third term 
might be singular. If we introduce the new coordinate $\ell=\sqrt{-2\rho}$, 
the second and the third term can be rewritten in $\ell \to 0$ as
\begin{align*}
 d\ell^2 + \ell^2(d\theta^2+\sin^2\theta d\vp^2).
\end{align*}
This metric is smooth in $\ell \to 0$.

\begin{figure}
\begin{center}
  \includegraphics[width=4cm]{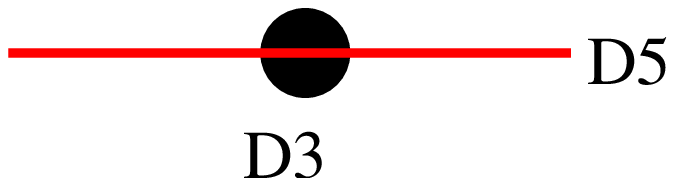} \qquad $\Longrightarrow$ \qquad
  \includegraphics[width=4cm]{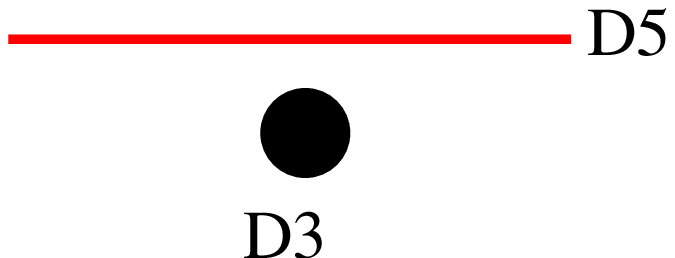}
\end{center}
\caption{The configuration of the D3-branes and the D5-brane. In the left 
 picture, 3-5 string have massless mode and the defect preserves the
 scale invariance in the gauge theory. In the deformation we done here,
  the D5-brane and the D3-branes are separated, 
  the 3-5 string become massive, and the scale invariance is violated.}
\label{fig2}
\end{figure}

\begin{figure}
\begin{center}
  \includegraphics[width=5cm]{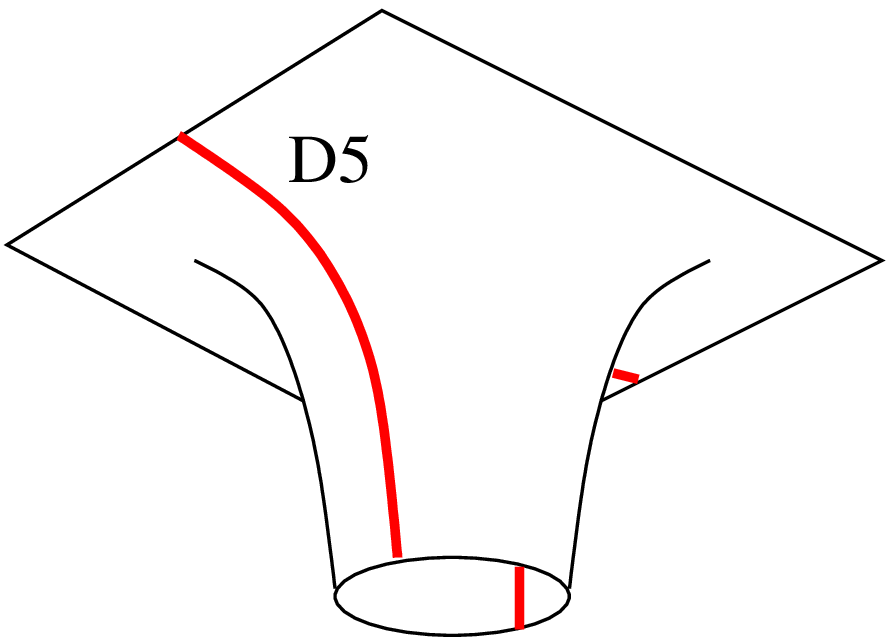} \qquad $\Longrightarrow$ \qquad
  \includegraphics[width=5cm]{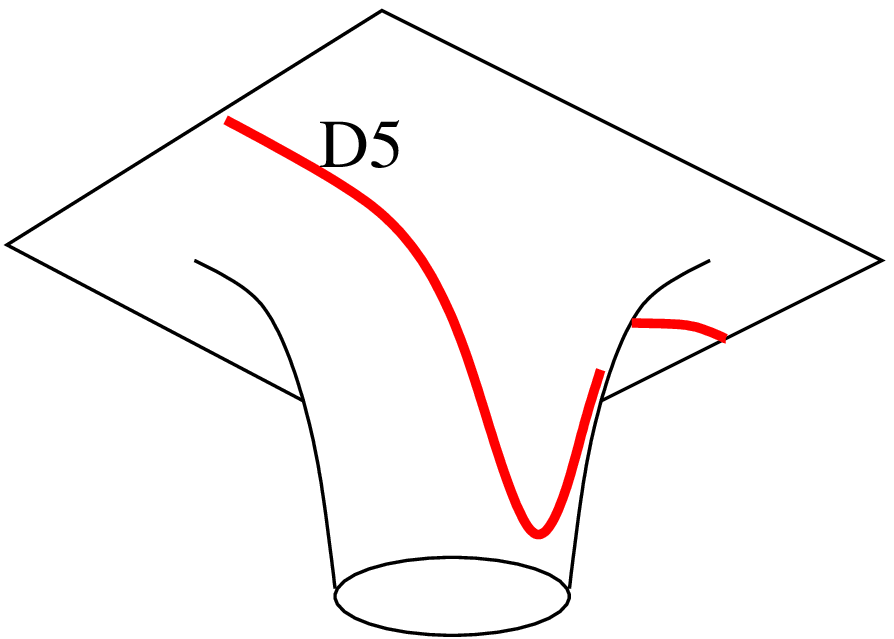}
\end{center}
\caption{Another picture of separating the D3-branes and the D5-brane. 
 The surface
 represents the geometry around the D3-branes and the bold line represents
 the D5-brane. If we add the mass to the fields on the defect, the D5-brane
 vanishes in the halfway of the throat.}
\label{fig3}
\end{figure}

Thirdly, the deformation we have done here is correspond to separating the
D3-branes and the D5-brane in the D3-D5 system as illustrated 
in figure \ref{fig2} and figure \ref{fig3}. If we separate the D3-branes and the 
D5-brane, the 3-5 string (the fields on the defect) becomes massive. 
Also, the D5-brane comes to stop in the halfway of the throat. 
These facts agrees with the solution we obtained here.

Finally, we expect this configuration preserves some supersymmetry since this 
configuration is the solution of the Bogomolnyi-like equation. It is also 
expected from the operator (\ref{mass-term}) in gauge theory side.
In the next subsection, we will see this fact in detail by considering
the Killing spinor and kappa symmetry projection on the brane.

\subsection{Supersymmetry of the solution}

Let us consider here the supersymmetry of the solution obtained in the previous 
subsection.
The surviving supersymmetry in the situation of D-branes embedded in 
supersymmetric vacuum is expressed by Killing spinors consistent with 
kappa symmetry projections of D-branes (see for example, \cite{Skenderis:2002vf}).
In this subsection, we set $R=g_s=1$ for simplicity.

First, we write down the Killing spinors of $AdS_5\times S^5$.
We use the vielbein of metric (\ref{metric}) 
\begin{align*}
 e^{a}=e^{-\rho}dx^a,\ (a=0,1,2,3),\qquad e^4=d\rho,\qquad e^5=d\psi,\qquad\\
e^6=\cos\psi d\theta_1,\qquad e^7=\cos\psi\sin\theta_1d\vp_1,\qquad
e^8=\sin\psi d\theta_2,\qquad e^9=\sin\psi\sin\theta_2d\vp_2.
\end{align*}
We also use the gamma matrices $\gamma^{A},\ A=0,\dots,9$ which satisfies
\begin{align*}
 \{\gamma^{A},\gamma^{B}\}=2\eta^{AB},\qquad \eta^{AB}={\rm 
 diag}(-1,1,\dots,1),\qquad A,B=0,1,\dots,9.
\end{align*}
In order to represent the two Majorana-Weyl spinor in IIB theory, we use one 
Weyl spinor $\epsilon=\epsilon_1+ i \epsilon_2$, where $\epsilon_1$ and 
$\epsilon_2$ are Majorana-Weyl spinor. The Killing spinor equation 
(invariance of gravitino) can be written as
\begin{align*}
 \left(\del_{A}
    +\frac12 \omega_{A}{}^{BC}\gamma_{BC}
    +\frac i2\gamma^{01234}\gamma_{A}\right)\epsilon=0,\qquad
         \del_{A}:=e_{A}^{M}\frac{\del}{\del x^{M}}.
\end{align*}
This equation can be solved as
\begin{align}
 \epsilon=e^{\frac12 \rho}\gamma_{4}
          e^{-\frac12 \gamma_{45} \psi} h \eta_{-}+
e^{-\frac12 \rho}e^{-\frac12 \gamma_{45} \psi }
            h\left(\eta_{+}+\sum_{a=0}^{3} x^{a}\gamma_{a}\eta_{-}\right),
\label{Killing-spinor}
\end{align}
where $h=h(\theta_1,\vp_1,\theta_2,\vp_2)$ is an unitary matrix dependent 
only on $\theta_1,\vp_1,\theta_2,\vp_2$. 
This $h$ satisfies the following differential equation.
\begin{align*}
 &\frac{\del h}{\del \theta_1}=-\frac12 \gamma_{46}h,
  &\frac{\del h}{\del \vp_1}=-\frac12
           \left(\gamma_6\cos\theta_1-\gamma_4\sin\theta_1\right)\gamma_7 h,\\
  &\frac{\del h}{\del \theta_2}=-\frac12 \gamma_{48}h,
  &\frac{\del h}{\del \vp_2}=-\frac12
           \left(\gamma_8\cos\theta_2-\gamma_4\sin\theta_1\right)\gamma_9 h.
\end{align*}
We need the concrete form of $h$ with $\theta_2=\vp_2=0$ (on the brane world 
volume) afterwards. This 
can be easily obtained as
\begin{align*}
 h(\theta_1,\vp_1,\theta_2=0,\vp_2=0)
   =\exp\left(-\frac12 
   \gamma_{46}\theta_1\right)\exp\left(\frac12\gamma_{67}\vp_1\right).
\end{align*}

In Eq.(\ref{Killing-spinor}), 
$\eta_{\pm}$ are constant spinors which satisfy
\begin{align*}
 i\gamma^{0123}\eta_{\pm}=\pm \eta_{\pm}.
\end{align*}
These constant complex spinors can be expressed by two Majorana-Weyl 
spinors $\lambda$ and $\chi$ as
\begin{align}
 \eta_{-}=\left(1-i\gamma^{0123}\right)\lambda,\qquad 
 \eta_{+}=\left(1+i\gamma^{0123}\right)\chi.\label{constant-spinor}
\end{align}
This shows that there are total 32 linearly independent Killing spinors.

Next, we consider the kappa symmetry projection. 
If we put some D-branes in this background, the surviving supersymmetry
corresponds to the Killing spinors consistent with kappa projection on the 
D-brane world volume
\begin{align}
 \Gamma\epsilon=\epsilon,\label{kappa-projection}
\end{align}
for the operator $\Gamma$ determined by the D-brane configuration.

In our solution obtained in the previous subsection, $\Gamma$ can be 
written as
\begin{align}
 \Gamma=-i\gamma_{0124}e^{\gamma_{45}\psi}\gamma_{67}K,\label{Gamma}
\end{align}
where $K$ is an anti-unitary operator which act on a spinor as 
$K\epsilon=\epsilon^{*}$ (charge conjugation) and commute with gamma matrices.
If we impose the condition (\ref{kappa-projection}) with (\ref{Gamma}) on the
 brane world volume to the Killing spinors 
(\ref{Killing-spinor}) with (\ref{constant-spinor}), we obtain the condition on
 $\lambda$ and $\chi$
\begin{align*}
 \lambda=0,\qquad \gamma_{3467}\chi=\chi.
\end{align*}
As a result, there are 8 linearly independent Killing spinors surviving in the 
solution obtained in the previous subsection.
This is the same amount as the supercharges of 3-dimensional 
$\Ncal=4$ super Poincare symmetry, and consistent to the picture in the gauge 
theory side. 

\section{\gf-theorem}

In this section, we consider the \gf-function and the \gf-theorem (analogue of
c-function and c-theorem in ambient field theory) in defect CFT.
First, we extend the definition of \gf-function to higher dimensional CFT.
Next, we define the \gf-function from the point of view of the holography, and
prove this \gf-function is monotonically non-increasing.

In this section, we consider $n$-dimensional ambient CFT and $(n-1)$-dimensional
defects. This means in string theory side $AdS_{n}$ brane in $AdS_{n+1}$.

\subsection{\gf-function in higher dimensional defect CFT}

Let us consider the $n$-dimensional CFT with $(n-1)$-dimensional defect at 
finite temperature. In high temperature and large volume,
the total free energy can be written as
\begin{align*}
 F=F_{\rm ambient}+F_{\rm defect}.
\end{align*}
The conformal symmetry implies the temperature dependences of $F_{\rm ambient}$ 
and $F_{\rm defect}$ are determined as
\begin{align}
 &F_{\rm ambient}=-av_{n-1}T^{n},\qquad
 F_{\rm defect}= -bv_{n-2}T^{n-1}, \label{free-energy}
\end{align}
where $T$ is the temperature, $v_{n-1}$ is the $(n-1)$-volume of the ambient 
space, $v_{n-2}$ is the $(n-2)$-volume of the defect, and $a$ and $b$ are 
dimensionless constants. Note that the coefficient $b$ represents the degrees of
freedom on the defect. By using this $b$, we define $g$ in higher dimensional 
CFT by the equation
\begin{align*}
 \ln g:=b.
\end{align*}
This is a natural extension of $n=2$ case \cite{Affleck:1991tk}
\footnote{
  In $n=2$ case, $v_0$ is interpreted as $1$, and 
$b$ is the defect entropy. That is $b=S_{\rm defect}
 :=-\del F_{\rm defect}/\del T$.}. 

Let us consider the RG flow from UV defect to IR defect in the 
same ambient. Philosophically, the degrees of freedom decreases along the RG 
flow. Thus, we expect the \gf-theorem: The \gf-function of UV defect is larger 
than that of IR defect.

\subsection{Holographic \gf-function}

In this subsection, we consider \gf-theorem from the point of view of the 
holography. The convenient metric of $AdS_{n+1}$ brane to treat the \gf-function 
is
\begin{align}
 ds^2=dr^2+\exp\left(-\frac{2r}{R}\right)\eta_{\mu\nu}x^{\mu}x^{\nu},\qquad 
 \mu,\nu=0,\dots,(n-1). \label{metric2}
\end{align}
The $AdS_{n}$ brane is sitting at $x^{n-1}=0$.

We propose a candidate of \gf-function in the string theory side as follows
\begin{align}
 \ln g =-\alpha T_{rr}.\label{g}
\end{align}
In above equation, $T_{rr}$ is the $rr$ element of the energy momentum tensor of 
the brane. $\alpha$ is an positive constant which might depend on
$R$ and parameters of the ambient theory. But $\alpha$ is 
independent of the energy scale and parameters of the defect theory.
Now, we will see this function has the following two properties
\begin{itemize}
 \item At the conformal point, this $\ln g$ agrees with the definition of the 
 previous subsection.
 \item $g$ is monotonically non-increasing.
\end{itemize}

Let us first consider the following simple example. In this model, a single 
scalar $\phi$ lives on the brane. We set the action of the brane as
\begin{align}
 \Ib=-\int d^{n}\xi \sqrt{-G}\left[\frac12G^{ab}\del_{a}\phi \del_{b}\phi
+V(\phi)\right], \label{scalar-action}
\end{align}
where we set $\xi^{j}=x^{j},\ j=0,\dots,(n-2),\ \xi^{n-1}=r$,  $G_{ab}$ is 
the induced metric of (\ref{metric2}), and $V(\phi)$ is the scalar potential.
The $rr$ element of the energy momentum tensor of this theory takes the value
\begin{align}
 T_{rr}=\frac12 \left(\frac{d\phi}{dr}\right)^2-V(\phi).\label{trr}
\end{align}

For each critical point $\phi_0,\ \frac{dV}{d\phi}(\phi_0)=0$, 
we obtain SO($n-1,2$) symmetric solution
$\phi(r)=\phi_0,$ (constant), where conformal symmetry is realized 
in the gauge theory side. The contribution of free energy from the defect is 
expressed as the brane action (in some regularization)
\begin{align*}
 F_{\rm defect}\propto \Ib=\int \sqrt{G} V(\phi_0)=V(\phi_0)\times 
 (\text{universal factor}),
\end{align*}
where the (universal factor) do not depend on $\phi_0$. Above equation shows
that the free energy of the defect and the factor $b$ in Eq.(\ref{free-energy})
is proportional to $V(\phi_0)$. On the other hand,
$T_{rr}=-V(\phi_0)$ is satisfied at conformal point. Therefore, 
$\ln g$ of Eq.(\ref{g}) is identical to $b$ in Eq.(\ref{free-energy}) at
conformal points if we set appropriately the constant $\alpha$ in Eq.(\ref{g}).

We can also show $T_{rr}$ in (\ref{trr}) is monotonically non-decreasing 
(that is, $g$ is monotonically non-increasing)
as a function of $r$. Actually, the $r$ derivative of $T_{rr}$ reads
\begin{align*}
 \frac{d}{dr}T_{rr}=\frac{d\phi}{dr}
          \left(\frac{d^2\phi}{dr^2}-\frac{dV}{d\phi}(\phi)\right)
          =(n-1)\left(\frac{d\phi}{dr}\right)^2\ge 0,
\end{align*}
where we use the equation of motion in the second equality.
 
This \gf-theorem can be proven not only in the model (\ref{scalar-action}),
but also in rather general system of brane in which
the conservation of energy-momentum and the weaker energy condition are 
satisfied. 
Actually, conservation of energy momentum and $(n-1)$ dimensional Poincare
invariance imply
\begin{align*}
 \frac{d}{dr} T_{rr}=\frac{n-1}{R}\left(T^{r}{}_{r}-T^{0}{}_{0}\right).
\end{align*}
The left-hand side of this equation is positive if we impose the weaker energy 
condition. Note that the  weaker energy condition is also essential to prove 
c-theorem in the holography\cite{Freedman:1999gp}.

Let us see the \gf-function of the flow obtained
in section \ref{section-solution}. By using the relation $R \rho=r$,
the $rr$ element of the energy momentum tensor reads
\begin{align*}
 T_{rr}=-4\pi R^2T_5 \cos^2 \psi
   \left[1+R^2 \left(\frac{d\psi}{dr}\right)^2\right]^{-1/2}.
\end{align*}
If we insert the solution (\ref{solution}), $T_{rr}$ becomes
\begin{align*}
 T_{rr}=-4\pi R^2T_5 \cos^3 \psi,\qquad \psi=\arcsin \exp(r/R).
\end{align*}
This $T_{rr}$ is actually monotonically increasing as the function of $r$.

\section{Conclusion}

In this paper, we investigated the RG flow on the defect from the point of view 
of the holography. 

First, we construct the solution of the D5-brane action which 
represents the RG flow of relevant deformation. We check that this deformation 
preserves right amount of supersymmetry. 

Next, we propose a candidate of \gf-function in brane theory. We prove 
that this \gf-function is monotonically non-increasing 
along the RG-flow by using 
the weaker energy condition. It is also checked that the \gf-function of the 
solution we obtained in this paper is actually monotonically non-increasing. 

One of the most important future problem is to find a flow from CFT to CFT. 
One simple realization of this kind of flow is adding flavour to the defect 
fields. This correspond to preparing the multiple D5-branes. Another way is 
to consider the defect connecting two different CFTs. This correspond to 
considering bent D5-branes \cite{Karch:2000gx}. 

It is also an interesting problem to consider the deformation of $AdS_2$ brane
in $AdS_3$. The 
corresponding defect CFT becomes two dimensional one (with one dimensional 
defect) and close to the setting of the original \gf-theorem. 
In this $AdS_3$ case, the string worldsheet approach is also useful 
\cite{Bachas:2000fr,Giveon:2001uq,Lee:2001xe,Hikida:2001yi,Rajaraman:2001cr,Parnachev:2001gw,Lee:2001gh,Ponsot:2001gt,Hikida:2002fh}. 
This makes it possible to analyse the correspondence 
including the stringy corrections.

%%%%%%%%%%%%%%%%%%% acknowledgement %%%%%%%%%%%%%%
\subsection*{Acknowledgement}
I would like to thank Tohru Eguchi, Yasuaki Hikida, Yosuke Imamura, Yuji 
Sugawara and Tadashi Takayanagi for useful discussions and comments.

This work is supported in part by Soryushi Shogakukai.

%%%%%%%%%%%%%%%%%%%% bibliography %%%%%%%%%%%%%%
\providecommand{\href}[2]{#2}\begingroup\raggedright\endgroup
\end{document}